\documentclass[acmlarge]{acmart}

\usepackage{booktabs} 
\usepackage{subfiles}
\usepackage{multirow}
\usepackage{wrapfig}
\usepackage{color}
\usepackage{hyperref}
\usepackage[inline]{enumitem}
\usepackage[ruled]{algorithm2e} 
\usepackage{mwe}

\SetAlFnt{\small}
\SetAlCapFnt{\small}
\SetAlCapNameFnt{\small}
\SetAlCapHSkip{0pt}
\IncMargin{-\parindent}

\acmJournal{IMWUT}
\acmVolume{0}
\acmNumber{0}
\acmArticle{}
\acmYear{2018}
\acmMonth{0}
\acmArticleSeq{0}
\acmPrice{15.00}

\setcopyright{acmlicensed}

\acmDOI{10.1145/3214261}

\received{November 2017}
\received[revised]{February 2018}
\received[accepted]{April 2018}

\begin{document}
\title[A Survey of Attention Management Systems in Ubiquitous Computing Environments]{A Survey of Attention Management Systems in Ubiquitous Computing Environments}  

\author{Christoph Anderson}
\orcid{0000-0002-4082-8457}
\affiliation{%
  \institution{University of Kassel}
  \department{Chair for Communication Technology}
  \streetaddress{Wilhelmsh{\"o}her Allee 73}
  \city{Kassel}
  \state{Hessen}
  \postcode{34121}
  \country{Germany}}
   
\author{Isabel H{\"u}bener}
\affiliation{%
  \institution{University of Kassel}
  \department{Chair for Communication Technology}
  \streetaddress{Wilhelmsh{\"o}her Allee 73}
  \city{Kassel}
  \state{Hessen}
  \postcode{34121}
  \country{Germany}}
  
\author{Ann-Kathrin Seipp}
\affiliation{%
  \institution{University of Kassel}
  \department{Chair of Business Psychology}
  \streetaddress{Pfannkuchstra{\ss}e 1}
  \city{Kassel}
  \state{Hessen}
  \postcode{34121}
  \country{Germany}}
  
\author{Sandra Ohly}
\affiliation{%
  \institution{University of Kassel}
  \department{Chair of Business Psychology}
  \streetaddress{Pfannkuchstra{\ss}e 1}
  \city{Kassel}
  \state{Hessen}
  \postcode{34121}
  \country{Germany}}
  
\author{Klaus David}
\affiliation{%
  \institution{University of Kassel}
  \department{Chair for Communication Technology}
  \streetaddress{Wilhelmsh{\"o}her Allee 73}
  \city{Kassel}
  \state{Hessen}
  \postcode{34121}
  \country{Germany}}
  
\author{Veljko Pejovic}
\affiliation{%
 \institution{University of Ljubljana}
 \department{Faculty of Computer and Information Science}
 \streetaddress{Vecna pot 113}
 \city{Ljubljana} 
 \country{Slovenia}}

\begin{abstract}
Today's information and communication devices provide always-on connectivity, instant access to an endless repository of information, and represent the most direct point of contact to almost any person in the world. Despite these advantages, devices such as smartphones or personal computers lead to the phenomenon of attention fragmentation, continuously interrupting individuals' activities and tasks with notifications. Attention management systems aim to provide active support in such scenarios, managing interruptions, for example, by postponing notifications to opportune moments for information delivery. In this article, we review attention management system research with a particular focus on ubiquitous computing environments. We first examine cognitive theories of attention and extract guidelines for practical attention management systems. Mathematical models of human attention are at the core of these systems, and in this article, we review sensing and machine learning techniques that make such models possible. We then discuss design challenges towards the implementation of such systems, and finally, we investigate future directions in this area, paving the way for new approaches and systems supporting users in their attention management.
\end{abstract}

%
%

%
%


\thanks{The authors would like to thank the participants and the organizers of Dagstuhl Seminar 17161 on Ambient Notification Environments for fruitful discussions that helped to shape our research vision. The
authors would also like to thank the anonymous reviewers for their helpful and constructive comments that greatly helped to improve the final version of the paper. This work has been co-funded by the Social Link Project within the LOEWE Program of Excellence in
Research, Hessen, Germany.}
\authorsaddresses{Authors' addresses: C. Anderson, I. H{\"u}bener, {and} K. David, Chair for Communication Technology, University of Kassel, Wilhelmsh{\"o}her Allee 73, 34121, Kassel, Germany; email: \href{mailto:comtec@uni-kassel.de}{comtec@uni-kassel.de}; Ann-Kathrin Seipp, and Sandra Ohly, Chair of Business Psychology, University of Kassel, Pfannkuchstra{\ss}e 1, 34121, Kassel, Germany; email: \href{mailto:ohly@uni-kassel.de}{ohly@uni-kassel.de}; V. Pejovic, Faculty of Computer and Information Science, University of Ljubljana, Vecna pot 113, Ljubljana 1000, Slovenia; email: \href{mailto:Veljko.Pejovic@fri.uni-lj.si}{Veljko.Pejovic@fri.uni-lj.si}}

\maketitle

\renewcommand{\shortauthors}{C. Anderson et al.}

\section{Introduction}
The emergence of computing and communication devices has radically changed the way we communicate and exchange information. It has been estimated that by 2025, the number of unique mobile subscribers will reach $5.9$ billion, representing over two-thirds of the world's population. Furthermore, the increased penetration rate of the mobile Internet -- over $5$ billion estimated by 2025 -- in combination with technological advancements of mobile devices, provides information access anywhere and anytime~\cite{Association:2018}. Such devices come with applications that help us navigate in unknown areas, keep distant social contacts alive, schedule our time, do shopping on the move, and in general help us with tasks that, in the past, were impossible or cumbersome to achieve.

Mobile computing and connectivity, in addition to anywhere-anytime access, provide a means for proactive information delivery. For example, online social networking applications can keep us up-to-date with other users' reactions to our posts; colleagues can inform us about work projects; retailers can send us discount offers as we are passing by shops. This kind of interaction provides seamless support for multitasking and better time utilization. Through mobile notifications, we can receive actionable information~\textit{in situ}. For example, an alert about a phone's battery status reminds us to charge it before leaving home; information about a sudden traffic jam down the road helps us to adapt our driving route; and a text message stating that a person we are scheduled to meet with will be $20$ minutes late allows us to use the extra time more efficiently. Nevertheless, such an opportunity to guide a user towards efficient multitasking often results in fragmenting a user's attention, reducing work performance~\cite{Leroy:2009}, increasing task error rates~\cite{Bailey:2006}, inducing stress~\cite{Mark:2008}, and even facilitating uninstalls of applications due to annoyed users~\cite{Pielot:2017a}. With the increasing number of applications and devices we use, the chances for irrelevant and unwanted interruptions rise. Consequently, an average mobile user is faced with about $100$ mobile notifications per day, many of which turn out to be irrelevant or distracting~\cite{Mehrotra:2015}. To ameliorate this issue, users may even completely disable all notifications, essentially rendering the whole mechanism designed to support multitasking irrelevant.
The decision on when to interrupt a user is left to application developers, and therefore, it is crucial to ensure that ubiquitous computing naturally supports, rather than interferes with, our everyday life when designing and developing intelligent attention management systems. Despite the fact that today's mobile applications take a rather \textit{ad hoc} approach to sending notifications, attention management has gained significant research traction among the ubiquitous computing community. 
A number of recent survey articles investigated attention management systems indicating the rising importance of this field. The existing reviews incorporated concepts from cognitive sciences~\cite{Roda:2006}, or focused on technical realizations of anticipatory mobile computing, including interruptions as part of personalized interactions~\cite{Pejovic:2015}. The survey of interruptibility prediction in ubiquitous systems by Turner et al. is the one most related to our work. Yet it focuses on a meta-analysis of selected interruptibility management approaches and examines specific aspects of these, namely scenario selection (e.g., lab vs. in-the-wild), data collection, and prediction of interruptibility through machine learning~\cite{Turner:2015}.
In this article, we rather expand the investigation of attention management systems, following a~\textit{combined} approach by providing \begin{enumerate*}[label=(\roman*)]\item{a holistic investigation of the theoretical underpinnings of human attention},\item{an overview of the existing approaches to attention management in ubiquitous computing},\item{guidelines towards designing attention management systems},\item{recommendations for future research in this field, so to close the gap between the cognitive theory and ubiquitous computing practice}\end{enumerate*}. 

This article is organized as follows: In Section~\ref{sec:definition}, we first examine definitions of the term~\textit{attention} and explore factors that capture it. We further discuss how interruptions in ubiquitous computing environments are created and handled. In Section~\ref{sec:theory}, we present recent theories on how the human brain processes interruptions. Our focus is to bring experiences from cognitive psychology closer to ubiquitous computing developers and is driven by the observation that attention management systems need to~\textit{sense} and~\textit{understand} the user to enable adaptive attention management. Equipped with sensors that provide a view into a user's contextual environment, and to a certain extent, even into an individual's internal state (e.g., stress~\cite{Lu:2012}, emotions~\cite{Rachuri:2010}), modern mobile devices are well suited for this purpose. Section~\ref{sec:ams} provides a detailed review of approaches towards sensing and modeling users' behaviors concerning mobile interruptions. We examine different sensing modalities and machine learning approaches used for interruptibility inference, as well as metrics towards evaluating interruptibility models. Finally, in Section~\ref{sec:implications} we examine the implications that different approaches of interruption management can have on our health and well-being. Further, we identify discrepancies between the rules that govern our cognitive processes and the current state-of-the-art systems for attention management in ubiquitous computing, such as to interrupt at task boundaries, allow users to rehearse tasks, or provide hints to help users return to the original task after an interruption, setting guidelines for future research in this area.\\

\noindent\fbox{%
	\parbox{.98\textwidth}{%
		\begin{small}
			\textbf{Key takeaway points}
			\begin{itemize}
				\item[$\star$] {Attention is captured and steered by external and internal stimuli, while stimulus properties, such as its duration, location, intensity, etc., impact the user's reaction to the stimulus. Multimodal alert types, e.g., sound, light, vibration, and multiple device environments call for a coordinated judicious use of alerting in ubiquitous computing.}
				\item[$\star$] {Limited cognitive capacities and threaded task processing imply that, in order to minimize disruptions, interruptions need to arrive at task boundaries or during routine tasks, should allow for task state rehearsal, and support context retrieval through hints presented to the user after an interruption.}
				\item[$\star$] {Sensor data from ubiquitous computing devices reveals a user's location, physical activity, collocation with other people, and other information of a user's context that can be related to interruptibility. Next generation wearable devices and personalized machine learning models promise to bring us closer to direct inference of a user's cognitive processes.}
			\end{itemize}
		\end{small}
	}
}

\section{Attention and Interruption: Definitions and Strategies}
\label{sec:definition}
In this section, we review definitions of the terms~\textit{attention} and~\textit{interruption}. We examine the connection between attention shifting and interruptions and discuss how interruptions can be handled through attention management systems in ubiquitous environments.

\subsection{What is Attention?}
\label{sub:attention}
There is no common understanding of~\textit{attention} in the literature. Attention is often considered as selective processing of incoming sensory information~\cite{Driver:2001}, with limited capacity~\cite{Chun:2011} and reactive and deliberate processes~\cite{Posner:1982}. Attention is also referred to as the ability to ignore irrelevant information~\cite{Chun:2001}. The process of selecting stimuli can be voluntary or be steered by external events. The former type refers to goal-driven attention~\cite{Yantis:1998}, whereas the latter relates to stimulus-driven attention~\cite{Corbetta:2002}. In this article, we focus on stimulus-driven attention further referring to the definition of Ashcraft et al. who define attention as: \begin{quote}\textit{"the mental process of concentrating effort on a stimulus or mental event."}~\cite{Ashcraft:2006}\end{quote} Guided by these definitions, attention is considered as an internal cognitive process that allows an individual to select tasks or information that will be actively processed.

\subsection{External vs. Internal Stimuli}
\label{sub:stimuli}
Maintaining attention, for example, on a task is often difficult as we are repeatedly confronted with numerous stimuli that compete for our attention. A stimulus, once it captures our interest, leads to an attentional shift changing the selection from a previous stimulus to another. Attentional shifts can be induced by internal or external stimuli. For instance, motivation, thoughts, or emotions represent internal stimuli, e.g., a researcher's desire to work on a scientific paper. External stimuli are related to events originated from the surrounding environment, such as a ringing smartphone drawing our attention from working on a project proposal. Attentional shifts caused by internal stimuli are considered as internal interruptions, whereas external stimuli lead to external interruptions~\cite{Miyata:1986,Dabbish:2011}. This paper focuses on automated systems that manage external interruptions (e.g., notifications) supporting individuals in maintaining their attention on tasks and activities. In the following section, we describe the structure of interruptions and their effects on individuals, activities, and tasks.

\subsection{Interruptions and Distractions: Structure and Handling}
\label{sub:interruptions}
Interruptions are investigated in various research fields, e.g., the psychology of attention, or Human-Computer-Interaction (HCI). In fact, the term interruption is defined differently, according to the requirements of the respective research domain\footnote{We refer the reader to~\cite{McFarlane:1997} for a review.}. A particular definition in the field of psychology states that an interruption is~\begin{quote}
\textit{"an introduction to a new task or tasks on top of the ongoing activity, often unexpectedly, resulting in conflicts and loss of attention on the current activity, failing to resume the work where it was interrupted."}~\cite{Miyata:1986}
\end{quote} This definition implies that an interruption affects an individual's current activity, resulting in the loss of attention and the inability to resume the activity where it was interrupted. Interrupting a task means to abandon a current task before finalization and diverting the attention to a new or different task~\cite{Dabbish:2011}. The definition of interruption is closely related to distraction, which is researched in the field of road safety (see~\cite{Regan:2011} for a review). The difference between distractions and interruptions is that the former are encountered stimuli intended to be ignored~\cite{Clapp:2012} and filtered out by a top-down suppression of signals from the prefrontal cortex~\cite{Chao:1995,Chao:1998}. The latter are stimuli that represent aspects of a secondary task, resulting in a reallocation of cognitive resources~\cite{Clapp:2012}. For the scope of this article, we refer to the term \textit{interruption}, further investigating the structure of interruptions as well as handling strategies and approaches.

\subsubsection*{Structure of Interruptions \& Handling Strategies}
\begin{figure}
    \centerline{\includegraphics[scale=0.5]{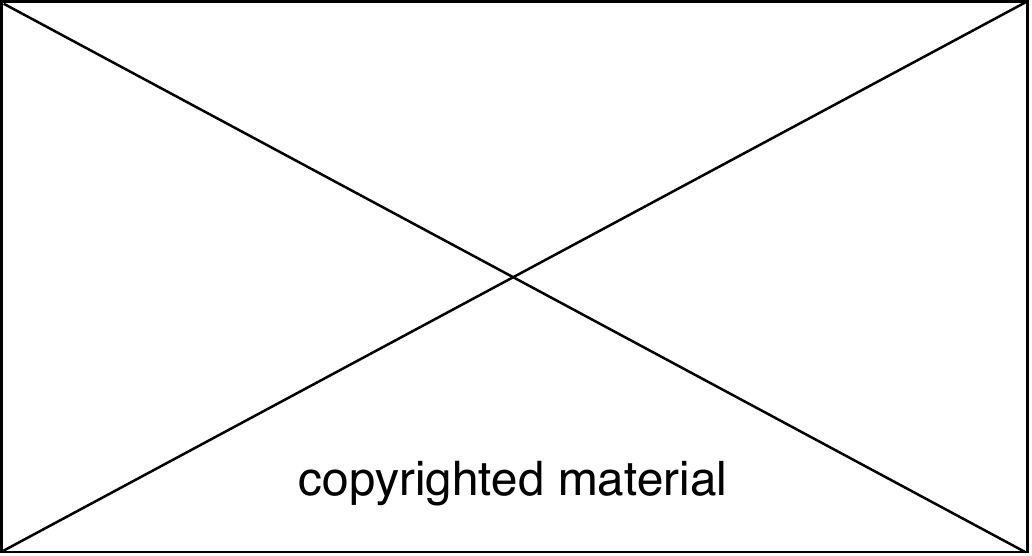}}
    \caption{Interruption model by Brixey et. al~\protect\cite{Brixey:2007}, \textcopyright 2007 Wolters Kluwer Health | Lippincott Williams \& Wilkins}
    \label{fig:interruption_model}
\end{figure}

Interruptions are not single events that interfere with an individual's current task or activity. Instead, they are embedded in a complex process that incorporates phases before, during, and after an interruption (see Figure~\ref{fig:interruption_model}). Before an interruption, an individual is focused on a task, known as the pre-interruption phase~\cite{Brixey:2007}, or primary task performance~\cite{Altmann:2004}. At a certain moment (end of Step $i$ in Figure~\ref{fig:interruption_model}) an individual perceives an interruption (e.g., phone call). This realization is followed by an~\textit{interruption lag}, a brief transitional period that precedes a pending interruption. In this period, an individual is aware but has not yet engaged the interruption, providing an opportunity to complete thoughts or to negotiate trending activities~\cite{Altmann:2004,Altmann2002}. For example, knowing that a colleague has already entered the office, a researcher might finish a sentence within a research paper first, rather than responding to the interruption immediately. After perceiving an interruption, an individual decides whether the interruption should be handled momentarily, later or not at all. Four general strategies for dealing with interruptions have been proposed in the literature~\cite{McFarlane:1997,McFarlane:2002a,Brixey:2007}. In particular, an individual might decide to stop working on a task to focus on the interruption~\textit{immediately}. Also, an individual can~\textit{negotiate} an interruption, for example, by first finishing a sentence of a research paper before responding to a colleague. Interruptions can also be~\textit{scheduled} or~\textit{mediated}. In the former case, an interruption is handled at a specific time or within a given period (e.g., every 15 minutes). In the latter case, an interruption is mediated by another person, proxy or system that interrupts an individual indirectly and decides~\textit{when} and~\textit{how} to interrupt. Before resuming, there is a lag between the interruption and the first action on the primary task. The~\textit{resumption lag} is a time span, where an individual rather needs to collect thoughts or hints about the last known status of the primary task before resuming to it~\cite{Altmann:2004}, even when an interruption was only perceived and not actively handled. These procedures are undertaken in the post-interruption phase, specifically in~\textit{step r} depicted in Figure~\ref{fig:interruption_model}.

\subsection{General Effects of Interruptions}
Most of the previous research agrees that interruptions result in prolonged completion times and increased error rates for \textit{primary tasks} that participants initially worked on~\cite{Bailey:2006}. For instance, Speier et al. analyzed the effect of interruptions in a laboratory study using computer-based tasks and concluded that both, the error rate and the completion time of the primary task, are negatively affected by an interruption, but only if the interrupting task is complex enough~\cite{Speier:1999}. Cutrell et al. state that the completion delay originates from switching from the primary task to the one signaled by the notification, as well as from switching back to the primary task~\cite{Cutrell:2000}. 

Regarding the effect on the secondary task, findings are less univocal. Bailey and Konstan's experiments show that secondary tasks are performed faster when delivered as an interruption to the primary task, rather than stand-alone tasks~\cite{Bailey:2006}. The authors deduce that this comes from a user's increased motivation to get back to and complete interrupted primary tasks as quickly as possible. At the same time, in cognitive load research, the reaction time, accuracy, and error rate on the secondary task are assumed to deteriorate, not improve, and the secondary task performance is often used as a measure of cognitive load imposed by the primary task~\cite{Paas:2003}. The subtle difference might arise from the way experiments are designed: interruptibility research often assumes that users are task-switching so that cognitive resources are focused on a single task at all times, while cognitive load research assumes that tasks are done in parallel so that resources are shared among multiple tasks.

\subsection{Interruptions in Ubiquitous Computing}
The evolution towards ubiquitous computing has drastically changed the way we communicate and receive notifications. Computing devices are now ubiquitous, personalized and provide always-on connectivity. Devices such as smartphones allow their owners to operate in parallel spheres by, for example, being in a phone call with one person, instant messaging with another while performing an unrelated task in the physical environment. Although, as early as in 1991 Mark Weiser started calling for \textit{"stealth"} ubiquitous computing, where devices quietly blend with the user's lifestyle~\cite{Weiser:1991}, as of today, we are still far from a realization of this vision. The rapid interweaving of ubiquitous computing in our everyday life facilitates numerous channels for communication while at the same time leaving only a few barriers for attention capturing. The net effect of uncontrolled exposure to interruptions is negative, especially in terms of task completion~\cite{Bailey:2006}, performance~\cite{Leroy:2009}, and emotional states~\cite{Mark:2008}.

Whereas the effect of interruptions on task performance was the primary focus of early research on interruptibility, understanding the impact mobile notifications have on a user's internal state emerged as a critical research question in ubiquitous computing. Increased stress and frustration caused by computer-based interruptions have been observed by Adamczyk and Bailey~\cite{Adamczyk:2004}. The authors found that the timing of interruptions within task execution has a significant effect on an individual's perceived amount of stress and frustration. More stress and frustration is induced when interruptions are timed between the execution of interrelating subtasks and less if timed after subtask completion\footnote{In Section~\ref{sec:theory}, we discuss the cognitive background of this finding in detail.}. A study by Mark et al. confirms stress-inducing effects of instant messaging and phone interruptions~\cite{Mark:2008}. Furthermore, Kushlev et al. investigate effects of interruptions caused by notifications on a group of $221$ participants who are exposed to the maximum level of mobile notifications for one week and for another week assigned to the minimum amount of such notifications~\cite{Kushlev:2016}. Their findings show a positive correlation between interruptions and symptoms of hyperactivity and inattention of individuals.

With ubiquitous connectivity, there is a need to be up-to-date with the events of interest. Internal interruptions (see Section~\ref{sub:stimuli}) are increasingly common and new behavior patterns, such as constant notification checking, have appeared. Kushlev and Dunn demonstrate the negative effect of frequent email checking on users' well-being~\cite{Kushlev:2015}. In another study about participants, who deliberately disabled all push notifications on their smartphones, Pielot and Rello found that the lack of notifications indeed increases participants' self-reported productivity, but at the same time leaves the participants anxious about not being responsive as expected and being less connected with their social contacts~\cite{Pielot:2017}. Finally, an earlier study of instant messaging users finds that without these messages users fallback to in-person interruptions which in certain work environments lead to even more stress~\cite{Garrett:2007}. These examples imply that ubiquitous computing does not present an inherently negative technological determinant. However, attention management systems need to be designed carefully to facilitate multitasking, minimize anxiety and the sense of social isolation, while ensuring that users are not overloaded with information and distracted in their work.

\section{Multitasking: How  Interruptions Are Processed}
\label{sec:theory}
A major issue with attention management in ubiquitous computing environments is the dissonance between interruptions, often signaled by mobile notifications, and the flow of cognitive engagement of users. The human brain, although sophisticated, is still limited when it comes to processing concurrent signals and tasks. In this section, we elaborate on the theoretical frameworks that explain how our brains handle tasks competing for the limited cognitive resources we have. We start with the unified theory of multitasking developed by Salvucci and Taatgen~\cite{Salvucci2008}, which is itself based on Anderson's Adaptive Control of Thought-Rational (ACT-R) architecture~\cite{Anderson1996}. This experimentally validated framework provides an analytic model of the human cognitive capacities and their use. We then examine how cognitive resources are used when individuals face concrete tasks and turn to the cognitive load theory to explain how interruptions interplay with ongoing tasks of varying complexity. The result of the analysis is the identification of the theoretically most suitable moments to interrupt individuals. Finally, we consider extensions to the presented models that indicate the importance of other factors, such as the content of the interrupting task and the social pressure on the user, or the way interruptions influence a user's reaction. These findings are further discussed in Section~\ref{sub:challenges_ams}, where we derive practical guidelines for designing attention management systems that are based on theoretical findings.

\subsection{Theory of Multitasking}
Fueled by the advances in information and communication technologies, the number of information and task sources keeps increasing, straining the limited cognitive resources we have. The understanding of cognitive processes is essential for efficient attention management in ubiquitous computing environments. In this article, we distill from the vast research in psychology and neurology that is concerned with multitasking and present the theoretical foundations of concurrent task processing. 

\subsubsection{Threaded Cognition} Threaded cognition is one of the most recent theories that explains multitasking via parallel threads competing for the same processing resources~\cite{Salvucci2008}. The concept is quite similar to the way multithreading is implemented in modern single-CPU/core computers: 
\begin{itemize}
	\item {Each active thread is associated with its own goal.}
	\item {Resources execute processes exclusively in service of one task thread at a time.}
	\item {Threads acquire and release resources in a greedy, polite manner.}
	\item {When multiple threads contend for the procedural resource, the thread with the highest urgency proceeds.}
\end{itemize}

\begin{figure}[ht]
	\centerline{\includegraphics[scale=0.5]{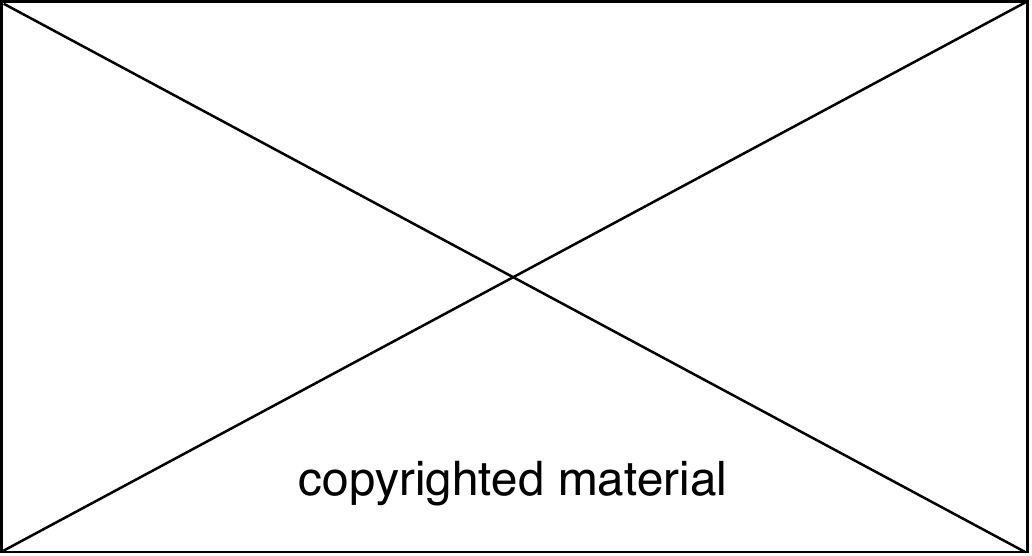}}
	\caption{Processing stream in threaded cognition~\protect\cite{Borst:2010}, \textcopyright 2010 American Psychological Association }
	\label{fig:borst_simple}
\end{figure}

The resources these threads compete for are cognitive resources, including declarative resources representing static knowledge that can be recalled (or forgotten), procedural resources representing procedural skills as goal-directed production rules, and perceptual and motor resources, allowing information acquisition from the environment and actions in the environment, respectively. Even the simplest tasks usually require more than a single kind of a resource, and complex patterns of interference may arise when two or more tasks are to be executed in parallel. Take this simple example from Borst et al.'s experiment~\cite{Borst:2010}, depicted in Figure~\ref{fig:borst_simple}. One task requires the user to press a key when a visual stimulus is shown (white boxes), while the other requires a vocal response from the user after an auditory stimulus (gray boxes). The x-axis represents time, and the boxes correspond to the period in which a resource is used (either by one task or the other). The tasks commence by activating a production rule related to a corresponding stimulus and continue with the encoding in either the visual or the aural module. We see that the interference (marked with \textbf{A}) caused by the contention for the procedural module prevents the perfect in-parallel execution of the task. Such interference can lead to delayed completion times and errors. 

\subsubsection{Computational Modeling of Multitasking} How exactly the brain harnesses cognitive resources, how much processing time each operation takes, and how the resources map to different parts of the brain is explained by cognitive architectures, for instance, the ACT-R architecture developed by Anderson~\cite{Anderson1996}. This theory describes the procedural memory regarding production rules, while declarative knowledge gets retrieved in chunks as necessary. Different modules, such as visual, aural, and manual, represent interfaces both for perceiving the physical world, as well as acting in it. However, the central production system is not sensitive to everything that happens within these modules. Instead, it can only respond to a limited amount of information that is deposited in the buffers of these modules. In other words, the ACT-R architecture puts a limit on human attention. Similarly, only the currently retrieved facts, and not all the information in our long-term memory, are readily available for the production system. In addition to explaining cognitive processes on the conceptual level, ACT-R also provides a computational tool, enabling fine-grain simulation of task duration\footnote{An open-source Lisp implementation of the framework is available at the following URL: \url{http://act-r.psy.cmu.edu/software/}}. This theory has been verified through a number of studies with different task types, from mathematical problems to driving simulations. Salvucci and Taatgen use ACT-R for the implementation of their threaded cognition theory and extend the computational model to support multitasking~\cite{Salvucci2008}. They confirm that the postulates of the threaded cognition theory realistically describe cognitive processes that happen when humans handle concurrent tasks. 

To understand how interruptions affect cognitive processes, we have to examine concepts associated with the primary and the interrupting task. According to the ACT-R theory, when recovering from an interruption, primary task chunks that were stored in declarative memory have to be retrieved. The easiness at which they are retrieved is related to their \begin{enumerate*}[label=(\roman*)]\item{base-level activation}, a measure of a chunk's inherent likelihood of being recalled, and \item{strength of association}, a measure of how the current environment context facilitates cueing of the particular chunk\end{enumerate*}. For attention management system designers, this presents a clear case for including hints about the interrupted task to facilitate task resumption. 

\begin{figure}[htbp]
	\centerline{\includegraphics[scale=0.5]{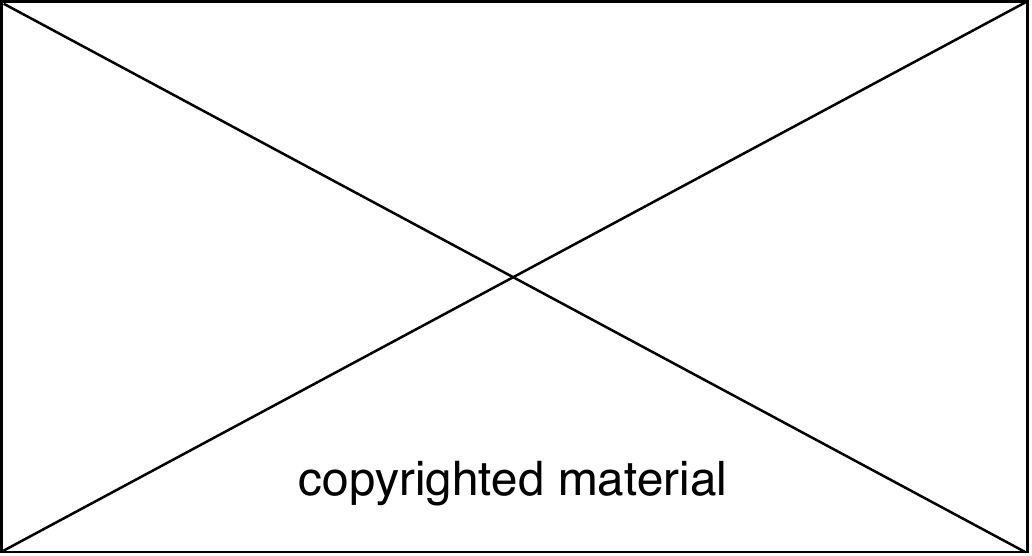}}
	\caption{ACT-R Architecture of cognitive resources~\protect\cite{Salvucci:2010a}, \textcopyright 2011 Dario Salvucci \& Niels Taatgen}
	\label{fig:cognitve_architecture}
\end{figure}

Following the ACT-R theory, each task has an associated goal that can also include subgoals. This is a cornerstone of Altmann et al.'s theory called Memory-for-Goals~\cite{Altmann2002}, according to which a task interruption results in the current active goal being stored in declarative memory. Task suspension and resumption processes can then be broken down into the core processes of rehearsal (or strengthening) and recall. When the task is suspended, the stored associated task goal decays, albeit its activation is raised with rehearsal. According to Altmann et al., the cognitive system must strike a balance between rehearsing the suspended task so that it can be easily retrieved when needed, but at the same time not activate it to the point where it interferes with the newly-attended task. The authors suggest that goals are preferably retrieved when they are associated with a cue, and when the goal-cue relationship is built just before the goal was suspended. 

The Memory-for-Goal theory was also used to explain the difference in the time needed to revert to the primary task after an interruption depending on the interrupting task duration and complexity. In the experiment by Monk et al., users performed a complex Video-Cassette-Recorder programming task, while getting interrupted with tasks of different complexity~\cite{Monk2008}. Resumption lags were longer for complex tasks than for simple tasks. Also, longer interruptions resulted in longer task resumption. This can be explained by the time decay of goal activation proposed in~\cite{Altmann2002}. Monk et al. discuss immediate design guidelines grounded based on these results, such as an in-car navigation system that should strive to minimize the interruption complexity, not necessarily the interruption duration, as simple interruptions lead to relatively fast primary task resumption, irrespective of the duration. However, neither of the theories investigate nuances in interruption complexities or why the resumption lag does not seem to scale to a certain level with interruption complexity.

\subsubsection{Memory for Problem State} The explanation of why complex tasks cause stronger interferences is provided by Borst et al.~\cite{Borst:2015}. The authors build upon the threaded cognition theory. Threaded cognition allows for multiple parallel goals, and thus multiple tasks (threads), to be active. This translates into the assumption that the goal module in ACT-R can represent several goals at the same time. Depending on its complexity, a task may or may not require \textit{the problem state}. The problem state resource is used to maintain intermediate mental representations that are necessary for performing a task. Borst et al. give an example of an algebra problem, such as $2x-5=8$, where the problem state can be used to store the intermediate solution, i.e., $2x=13$. There is also physiological evidence of the existence of the problem state, as neuroimaging experiments find that the transformation of mental representations correlates with the blood oxygen level-dependent activity in the posterior parietal cortex~\cite{Anderson2005,Sohn:2005}. At the moment of interruption, if both the primary and the interrupting task are complex, the problem state of the primary task is stored in declarative memory where it starts to decay. On the return from the interruption, the primary task's problem state has to be retrieved from the memory, yet, this process is tied to delay and errors. Even more delays and errors are to be expected if the problem state has lingered in the declarative memory for a longer period of time. Experimental results with tasks of differing complexities confirm this theory and show that when either the primary or the interrupting task is simple enough not to require the problem state, the resumption time and the task errors are minimized. 

\subsection{Beyond Interruption Timing}
The theories outlined in this section are bound to have substantial implications on how attention management systems might evolve (see Section~\ref{sub:challenges_ams}). However, these theories do not present a comprehensive explanation of interruptibility. This is particularly true for attention management in ubiquitous computing environments where other aspects, such as the type of the communication channel, the content of the interruption/message, the frequency of interruption, the relationship between the sender and the receiver, the alert type, and the social environment, may play a deciding role in how an interruption is handled. 

Considering the \textit{interruption content}, Speier et al. show that interruptions containing information dissimilar from the primary task take longer to complete than those with content related to the primary task~\cite{Speier:1999}. Since the success of the retrieval from declarative memory depends on the context, the closer the sought for information is to the currently active one, the easier it is to retrieve it. Addas and Pinsonneault consider interruptions that are either relevant to the task at hand, or not, and that are either actionable or informative~\cite{Addas:2015}. An actionable irrelevant interruption leads to task errors and longer completion times. Relevant information, on the other hand, helps with the task quality but may affect the time to finish the task. The findings are important for project management, for example. If one's task responsibilities are defined too narrowly, many interruptions that are not directly related to the task will be perceived as interruptions. Expanding the task boundaries allows for a more fluid definition of relevant interruptions. 

Grandhi and Jones propose a theoretical framework that explains interruption handling decisions, and takes into account the user's \textit{cognitive}, \textit{social} (the place the interruptee is in, the social aspect of the place, and people present at that place) and \textit{relational context} (who is the message from, what is it about, and the history of the interruptee-interrupter relationship)~\cite{Grandhi:2009}. Through experimental validations, the authors show the multifaceted nature of interruption handling. Even though the effects of interruptions on task performance or appropriateness in a social setting may be negative, individuals may still decide to handle an interruption. The authors argue that people may consider not just how an interruption affects one's local task, but also how it may affect other tasks and interpersonal relations. Consequently, the authors claim that instead of an automated interruption management system, efforts should be turned to User-Interface (UI) designs that convey the necessary relational context to the interuptee and facilitate one's interruption handling decision process. 

Personal devices such as smartphones or personal computers allow experimental validation and further empirical investigation of factors related to interruptibility. A large-scale study of mobile notification usage by Shirazi et al. shows that users assign different importance to notifications triggered by different application categories~\cite{Shirazi:2014}. This is unlikely to be related to the specifics of the application's notification management but is probably conditioned on the application design and the social role that a user assumes in different applications (e.g., Skype for business contacts, Whatsapp for family chats). That mobile computing users indeed handle interruptions from different contact types (e.g., family, friends, work) differently has been shown in a study by Mehrotra et al.~\cite{Mehrotra:2016a}. Pielot et al. found that users opt for immediate handling of potentially disruptive interruptions just to conform to social expectations, demonstrating the social pressure in notification handling~\cite{Pielot:2014}. 

Finally, physical properties of stimuli determine whether and how the attention is going to be steered. Two independently conducted studies have confirmed the difference in interruption handling times depending on whether a notification vibrates, blinks or sounds an alert~\cite{Pielot:2014,Mehrotra:2016}

\section{Attention Management Systems}
\label{sec:ams}
In this section, we examine functional principles of attention management systems, in particular, systems that (partly) seize the theories and frameworks mentioned before. According to Bailey and Konstan, an attention management system is defined as
\begin{quote}
\textit{" a system that computationally seeks to balance a user's need for minimal disruption and the application's need to efficiently deliver information."}~\cite{Bailey:2006}
\end{quote}
In other terms, an attention management system might aim to steer an individual's attention in case of an interruption by choosing a suitable output modality gently or even postponing interruptions to a time where it could be handled by a user. For example, if we consider push notifications as interruptions, this means that a system automatically decides \textit{when} and in \textit{which ways} an individual is notified. How does a system differentiate between disruptive and non-disruptive interruptions? The answer lies in a computational model that holds characteristic information about potential disruptive interruptions. In the following, we examine implementations of underlying interruptibility models as well as designs of practical attention management systems.

\subsection{Designing Attention Management Systems}
\begin{figure}
	\includegraphics[scale=0.45]{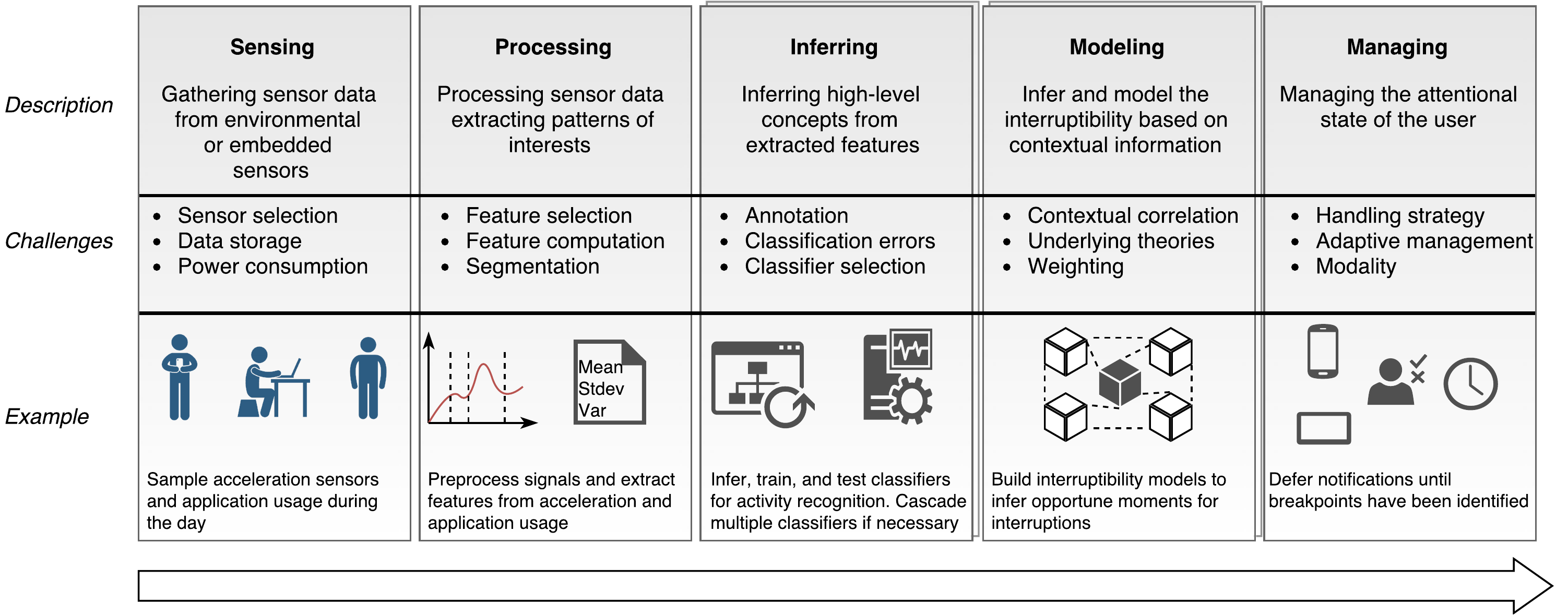}
	\centering
	\caption{Stages of an attention management system including sensing, processing, inferring, modeling, and managing. Key challenges along with an example are depicted for each stage.}
	\label{fig:stages}
\end{figure}

Sensing, actuating, and machine learning are at the core of attention management systems. These concepts interplay through different processing stages that connect raw data from the physical environment, over interruptibility modeling to attention management. These stages include \textit{sensing}, \textit{processing}, \textit{inferring}, \textit{modeling}, and finally \textit{managing} interruptibility as illustrated in Figure~\ref{fig:stages}. To demonstrate key challenges and implementation issues, take a running example of an attention management system inspired by Attelia II~\cite{Okoshi:2015a}, depicted in Figure~\ref{fig:stages}. Attelia II monitors acceleration signals and application usage to infer posture and locomotive activities of users, e.g., walking, sitting, and standing as well as a user's device usage. The system builds an interruptibility model that is based on theoretical postulates (see Section~\ref{sub:challenges_ams}) assuming that times between tasks represent opportune moments for interruptions. Finally, the system detects natural breakpoints~\cite{Newtson:1976} in physical and user-interface-based activities and postpones notifications to these moments.

\subsubsection{Sensors and Features}
\label{sub:features}
\begin{figure}[ht]
	\includegraphics[scale=0.5]{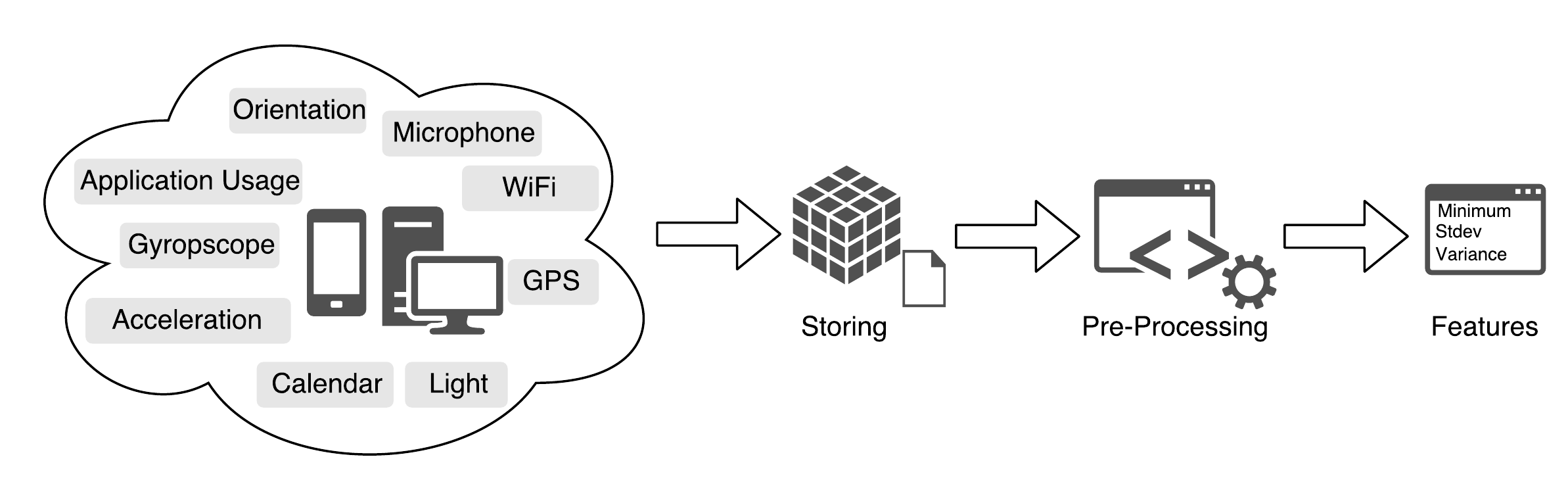}
	\centering
	\caption{Sensing and feature computation: From sensors to features}
	\label{fig:sensing}
\end{figure}

Equipped with microphones, light, acceleration, temperature, and other sensors, as well as communications interfaces, mobile computing devices can~\textit{perceive} their surrounding environment with the help of machine learning models. The key premise for building meaningful models is to select features, i.e., sampled data representations, that correlate with the classes of interest~\cite{Domingos:2012}. In Figure~\ref{fig:sensing}, we sketch the process of deriving features from sensed data. Raw sensor data (e.g., acceleration) is recorded, stored and then preprocessed, for example, by applying segmentation, sanitization, and normalization to remove artifacts such as outliers, noise or missing data. Features that correlate with targeted classes are then computed on preprocessed data. In this regard, a selection of the most informative features for a particular scenario is required given the vast amount of possible features. In general, such a selection is based on features that have been found useful to discriminate patterns in the domain at hand\footnote{We refer the reader to Section 4.3 of~\cite{Pejovic:2015}, which summarizes the most commonly used features in different inference domains.}. Table~\ref{tab:features} provides an excerpt of commonly used features in attention management systems. The sensors and features listed here are not exhaustive and are intended to be a starting point for designers of attention management systems. 

Besides sensors embedded in mobile devices, attention management systems may also draw information from personal computers as well as from sophisticated psychophysiological sensors. Psychophysiological sensors are used to extract information correlating with an individual's cognitive state (e.g., mental workload), whereas mobile devices and personal computers are found to provide additional characteristics about an individual's interruptibility, for example, related to activities, or application usage. The process of extracting features from sensed data is exemplified by Attelia II which uses acceleration sensors to infer an individual's posture and locomotive activities~\cite{Okoshi:2015a}. In the first step, raw acceleration signals are recorded and stored on the device. The system then computes $22$ time and frequency-domain features -- e.g., acceleration mean, variance, or correlation -- frequently used in the field of activity recognition that corresponds with an individual's movement.

\begin{table}
	\centering
	\caption{\textbf{Sensors and features commonly used in attention management systems to infer interruptibility.}\label{tab:features}}{
		\begin{scriptsize}
			\begin{tabular}{p{.15\textwidth}p{.3\textwidth}p{.475\textwidth}}
				\toprule %
				\textbf{Domain} & \textbf{Sensors} & \textbf{Features} \\
				\toprule %
				\textit{Physical Activities} &
				Acceleration~\cite{Urh:2016,Ho:2005,Kern:2004,Sarker:2014,Kern:2006,Komuro:2017,Okoshi:2015a}  & mean, energy, entropy, correlation, variance, ratio \{minimum, maximum\}  \\ \cmidrule{3-3} &
				Orientation~\cite{Poppinga:2014,Turner:2015a} & mean pitch angle, mean roll angle, flat, upright \\ \cmidrule{3-3} &
				Gyroscope~\cite{Urh:2016} & mean crossing rate  \\

				\midrule

				\textit{Device Interaction} &
				Microphone~\cite{Fisher:2011,Kern:2006,Bernstein:2008,Pielot:2017a,Gjoreski:2016,Malkin:2006}  & mean, variance of fourier coefficients, signal power, zero crossings, cepstral coefficients, noise \{true, false\}, ratio \{voice, noise, silence\}, spectral \{centroid, diffusion\} \\  \cmidrule{3-3} &
				Application usage~\cite{Okoshi:2015a,Okoshi:2015,Mehrotra:2015,Zuger:2017,Pielot:2017a,Tanaka:2011,Fogarty:2005a} & application name, window changed \{state, content\}, interface statistics (e.g., min, mean, max), skype status, number of applications per time frame, \\  \cmidrule{3-3} &
				Proximity~\cite{Poppinga:2014,Pielot:2014a} & screen covered    \\  \cmidrule{3-3}  &
				Light~\cite{Mehrotra:2016,Turner:2015a,Gjoreski:2016} & intensity, ambient light \{dark, dim, light, bright\}  \\  \cmidrule{3-3}  &
				Volume~\cite{Pielot:2014a,Turner:2015a,Pielot:2017a,Okoshi:2017} & unknown, silent, vibration, or sound \\  \cmidrule{3-3}  &
				Mouse and Keyboard~\cite{Zuger:2017,Tanaka:2011,Fogarty:2005a,Iqbal:2008} & click events, \{moved, scrolled\} pixels, keystrokes, most common key events \{pressed, released, traversed\} \\  \cmidrule{3-3}  &
				Calendar~\cite{Zuger:2017,Urh:2016,Zulkernain:2010} & events \{free, meeting\} \\  \cmidrule{3-3}  &
				Battery~\cite{Turner:2015a,Turner:2017,Gjoreski:2016} & charging \{true, false\} \\
                
				\midrule

				\textit{Location}   &
				GPS~\cite{Fisher:2011,Sarker:2014,Yuan:2017,Mehrotra:2015,Yuan:2017,Gjoreski:2016}  &  longitude, latitude \\ \cmidrule{3-3}   &
				WiFi~\cite{Mehrotra:2015,Kern:2006,Gjoreski:2016} & connectivity, coarse location \\

				\midrule
				
			    \textit{Temporal}  &
				Time~\cite{Turner:2015a,Pielot:2017a,Fisher:2011,Okoshi:2017,Zulkernain:2010} & time of day \{morning, afternoon, evening, night\}, day of week, weekend \{true, false\} \\  \cmidrule{3-3}  &
				Events~\cite{Pielot:2014a,Yuan:2017,Iqbal:2008} & time since: notification \{received, viewed\}, screen \{on, off, covered, uncovered\}, time to \{react, complete\} \\
				
				\midrule

				\textit{Psychophysiological}  &
				Electroencephalography (EEG)~\cite{Zuger:2015,Haapalainen:2010,Fritz:2014}  &  brain wave frequency bands (e.g., $\alpha (8-12Hz)$, $\beta(12-30Hz)$, $\gamma(30-80Hz)$, $\delta(0-4Hz)$, $\theta(4-8Hz)$), variance of $\theta$ power, median of $\beta$ power, \{min, max\} attention, \{min, max\} meditation, fractions of wave bands \{ $\Delta(\alpha/\beta)$, $\Delta(\beta/\gamma)$, $\Delta(\gamma/\theta)$\}  \\ \cmidrule{3-3} &
				Electrodermalactivity (EDA)~\cite{Zuger:2015,Fritz:2014,Rajan:2016,Haapalainen:2010} & \{min, max\} peak amplitude, $\Delta$number of phasic peaks/min, $\Delta$mean phasic peak amplitude, $\Delta$mean skin conductance level, skewness, \{mean, variance, median\} galvanic skin response\\ \cmidrule{3-3} & Electrocardiograph (ECG)~\cite{Sarker:2014,Rajan:2016} & \{ratio low/high-frequency components\} of heart beat and heart beat frequency, mean, median, stdev, percentiles \{10th, 25th, 75th, 90th\} \\ \cmidrule{3-3} &
				Photoplethysmograph~\cite{Zuger:2015} & \{mean, sum, max,\} peak amplitude BVP, mean heart rate, percentage difference of interbeat interval\{20ms, 50ms\} \\ \cmidrule{3-3} &
				Eye tracker~\cite{Haapalainen:2010,Fritz:2014,Rajan:2016} & \{mean, median, stdev.\} saccade duration, \{mean, variance, median\} pupil diameter, number fixations/min, \{mean, median, stdev\} fixation duration \\ \cmidrule{3-3} &
				Respiration~\cite{Sarker:2014,Rajan:2016} & duration of inhalation/exhalation, mean, median, stdev  \\ \cmidrule{3-3} &
				Temperature~\cite{Haapalainen:2010,Rajan:2016} & \{mean, variance, median\} heat flux, temp. difference \{nose, neck\} \\ 
				\bottomrule
			\end{tabular}
		\end{scriptsize}}
\end{table}

\subsubsection{Contextual Information and Machine Learning Techniques}
\label{sub:contexts}
\begin{figure}[ht]
	\includegraphics[scale=0.5]{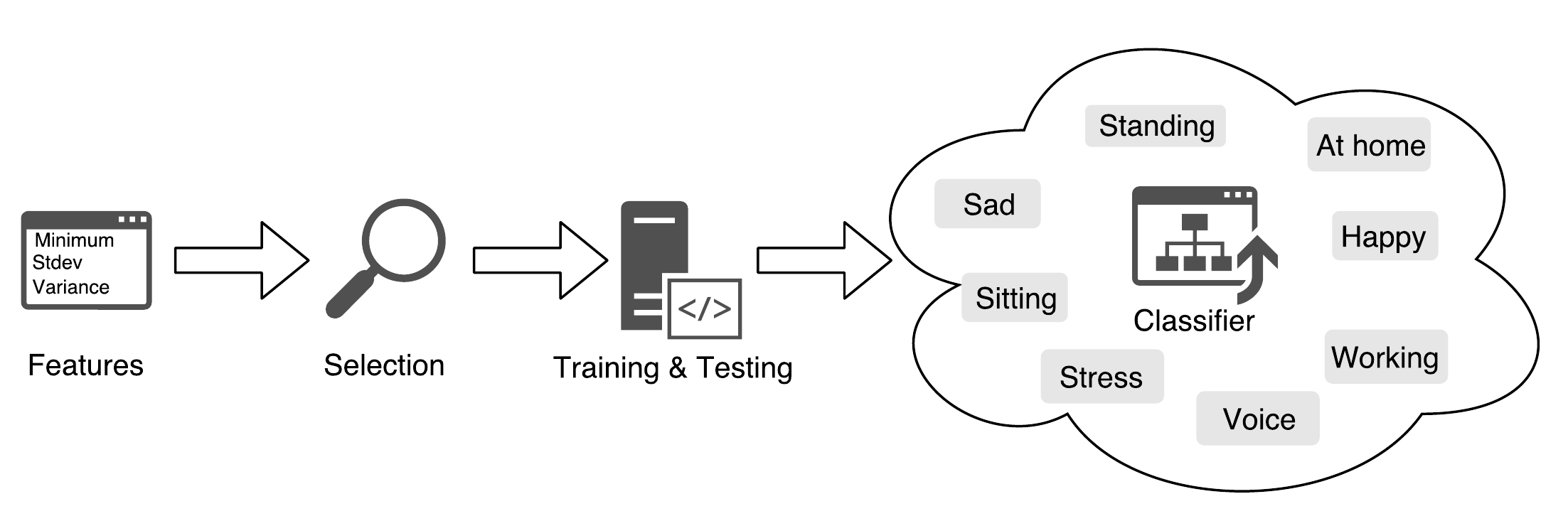}
	\centering
	\caption{From features to contexts: Deriving contextual information from sensor data.}
	\label{fig:context}
\end{figure}

The major theoretical determinants of interruptibility, such as the interplay of a stimulus signaling an interruption, a user's current task engagement, and the complexity of the interrupting and the interrupted tasks, were investigated in Section~\ref{sub:interruptions}. As attention management systems have often developed independently from underlying theories, they use contextual descriptors and features that reflect, \textit{but not necessarily determine}, the interruptibility state of an individual. In an \textit{optional} stage of inference, machine learning models use features to recognize contextual information of users and their environments, such as physical activities, semantic locations, or even emotions. Figure~\ref{fig:context} illustrates this process - features are first further pruned and compressed, for example, by applying the Principal Component Analysis~\cite{Wold:1987} or information gain~\cite{Hall:2003} to select a set of meaningful features. The resulting feature set is used to build machine learning models. As contextual information can be of different levels of abstraction, a plethora of contexts can be extracted through machine learning amplifying the possibilities of interruptibility modeling. A key challenge in the field of attention management, therefore, is to identify contextual descriptors that correlate with an individual's interruptibility, craft machine learning models that can infer these contexts, and finally tie them into an interruptibility model. 

Commonly used contextual information that correlate with an individual's interruptibility cover various aspects of life.\footnote{We refer an interested reader to the extended interruption taxonomy proposed in~\cite{Gievska:2005,Sykes:2014}.}. It is worth to note that not only physical activities~\cite{Ho:2005,Sarker:2014,Okoshi:2015a} or interactions~\cite{Poppinga:2014,Fisher:2011}, but also more expressive concepts such as complex activities~\cite{Kern:2006}, engagement levels~\cite{Pejovic:2015a}, even personal traits~\cite{Yuan:2017,Mark:2008} are frequently used in attention management systems. To infer such contextual information, various machine learning algorithms e.g., J48~\cite{Okoshi:2015a,Sarker:2014}, K-Means~\cite{Kern:2006}, or Support Vector Machines~\cite{Sarker:2014} are typically trained and tested\footnote{We refer the reader to~\cite{Bishop:2006} for an introduction to pattern recognition and machine learning.}. Given the vast amount of classifiers, a selection of classifiers depends on the amount and type of training data, on the processing power of the device, and above all on the classification performance. Statistical tests, e.g., Wilcoxon signed ranks or Friedmann tests, can help to evaluate and compare classification performances~\cite{Demsar:2006}.

\subsubsection{Interruptibility Models and Proxies}
\label{sub:models}
How to differentiate between disruptive and non-disruptive interruptions? The answer to this question lies in computational models which hold hypotheses about an individual's interruptibility. These models are either based on concepts and theoretical frameworks that have been constructed to explain the nature of human interruptibility (see Section~\ref{sec:theory}) or purely data-driven, based on the results of mining traces describing human behavior in different situations. Examples of the former are attention management systems that interrupt at task boundaries (breakpoints)~\cite{Adamczyk:2004,Iqbal:2005,Okoshi:2015a} or when task engagement is low~\cite{Pejovic:2015a}. These approaches are supported by underlying theories claiming that interruptions may require an exclusive resource -- the problem state -- which should not be occupied by the primary task at the time of the interruption~\cite{Borst:2015}. The latter, data-driven approaches, are realized by the abundance of sensor data gathered from mobile and personal computing devices. InterruptMe, for example, uses time-of-day information together with sensor readings from GPS and acceleration to build a model that explains user behavior concerning interruption handling~\cite{Pejovic:2014}. This approach is not based on any particular theory but might reveal links with existing theoretical frameworks after further investigations.

\begin{figure}[ht]
	\includegraphics[scale=0.6]{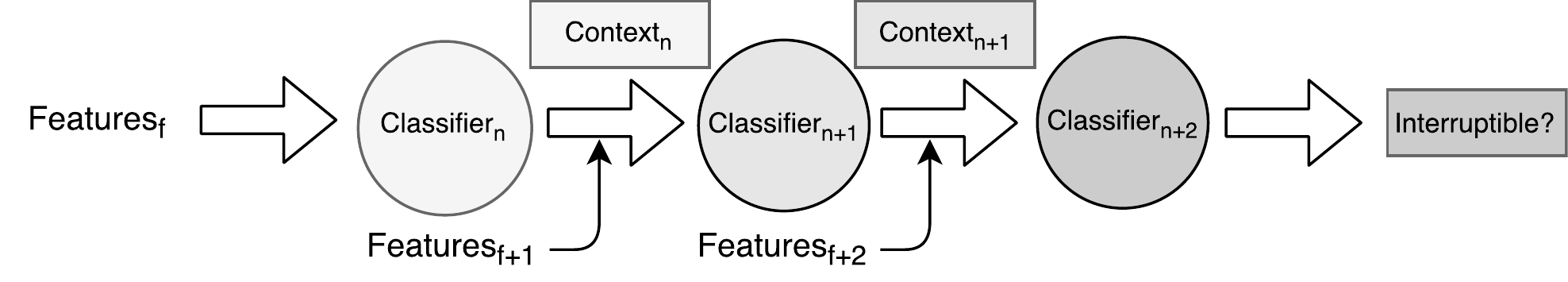}
	\centering
	\caption{Cascading classification: Combining multiple classifiers to infer interruptibility.}
	\label{fig:cascading}
\end{figure}

To construct interruptibility models, features carrying information sensed from the environment, can either be used directly (e.g., InterruptMe~\cite{Pejovic:2014}), or through further inference of contextual information (e.g., Attelia II~\cite{Okoshi:2015a}). This~\textit{cascading classification} -- a subsequent classification is trained on the output of a previous one -- facilitates different kinds of contexts, in particular, different levels of abstraction that augment interruptibility models (see Figure~\ref{fig:cascading}). For example, Attelia II classifies posture and locomotive activities and use this information to train a second classifier that infers breakpoints in physical activities. Combined with a breakpoint detection approach based on application usage, a final classification model is trained to infer interruptibility.

\begin{table}[ht]
	\centering
	\caption{\textbf{Interruptibility models and metrics of attention management systems.}
	\label{tab:models}}
	{\begin{scriptsize}
		\begin {tabular}{p{.13\textwidth}p{.2\textwidth}p{.275\textwidth}p{.3\textwidth}}
		\toprule %
		& \textbf{Model} & \textbf{Approach} & \textbf{Metric} 
		\\ \toprule
		Okoshi et al.~\cite{Okoshi:2017,Okoshi:2015a,Okoshi:2015,Okoshi:2014} &
		Breakpoints between physical activities/application usage   &
		Perceived mental workload is lowered by deferring interruptions to natural breakpoints in physical activities/application usage &
		Breakpoints via self-annotated application usage, mental workload via ESM: NASA-TLX~\protect\cite{Hart:1988}. Likelihood of physical breakpoints via ESM: $10$-Point Likert scale. Engagement to content via response time, click rate.
		\\ \midrule
		
		Pejovic et al.~\cite{Pejovic:2014} &
		Broader contextual information of individuals including activity, location, and engagement. &
		Interruptibility is reflected by the context in which an individual is at the moment of interruption. &
		Reaction presence: respond to a notification, time to reaction: respond to a notification within a time period, and sentiment towards notification via ESM: \{not at all, a little, some, very much\}.
		\\ \midrule
		
		Pejovic et al.~\cite{Pejovic:2015a} &
		Task engagement & 
		The degree to which an individual is engaged in a task determines interruptibility. &
		Levels of engagement: \{interesting, challenging, concentrated, and important\}, measure of engagement via ESM -  $4$-Point Likert scale: \{not at all, a little, somewhat, and very much\}. Questions: Is your current activity: \{interesting, challenging\}? How well are you concentrated on your task? Is the activity important for you?
		\\ \midrule
		
		Pielot et al.~\cite{Pielot:2014a} &
		Attentiveness   &
		The degree to which individuals pay attention to incoming notifications reflect interruptibility. Attending means that an individual either opens the application that received a notification or opens the notification drawer. &
		Attentiveness: \{very high, high, low\}, measured implicitly by the mobile phone usage.
		\\ \midrule
		
		Sarker et al.~\cite{Sarker:2014} &
		Availability    &
		Interruptibility is determined by an individual's capability of engaging tasks or activities. & Availability via EMA~\protect\cite{Shiffman:2008}, multiple times a day.
		\\ \midrule
		
		Yuan et al.~\cite{Yuan:2017} &
		Reaction \& Interruption intensity &
		Interruptibility is determined by first predicting an individual's reaction
		and second the intensity of an interruption. & 
		EMA~\protect\cite{Shiffman:2008} about participants' behaviors, physiological and psychological states, including $42$ items, current activity, and mood \{pleasant, unpleasant\} etc.
		\\ \midrule

		Z{\"u}ger et al.~\cite{Zuger:2015} &
		Mental workload &
		The perceived mental workload determines individuals' interruptibility. &
		Mental workload via ESM - $5$-Point Likert scale: \{(1) very-low, (5) very-high\}, perceived disturbance of interruptions: \{(1) not all disturbing, (5) very disturbing\}, interruptibility level: \{(1) highly 		interruptible, (5) not all interruptible\} Mental workload measured with psychophysiological sensors (see Table~\ref{tab:features}).	
		\\ \bottomrule
		\end {tabular}
		\end{scriptsize}}
	\end {table}
	
In Table~\ref{tab:models}, we list a selection of interruptibility models that are implemented in attention management systems. For example, an individual's engagement in tasks is investigated in~\cite{Pejovic:2015a} to predict interruptibility. The authors argue that individuals skilled in the task that is interrupted, feel less likely disturbed by an interruption. A possible explanation is that skilled individuals need less time to reconstruct the primary task state after an interruption due to stronger links in declarative memory. However, interruptions, during tasks that require a significant amount of concentration, are more likely perceived as disruptions. The reason for this might be that interruptions within periods of high mental workload -- induced by inexperience or difficult tasks -- result in higher resumption costs~\cite{Borst:2015} forcing an individual to spend more cognitive resources to resume to the primary task. The connection between mental workload and disruptiveness of interruptions has also been shown in~\cite{Iqbal:2005,Iqbal:2006,Cheng:2015}.

Different metrics and annotation techniques to measure an individual's interruptibility can be applied to evaluate respective models. For example, explicit measures include~\textit{experience sampling methods/ecological momentary assessments} (ESM/EMA) where individuals are asked to rate their current experience~\textit{in-situ}. In this regard, the NASA-TLX is frequently used, in particular, to assess models that are based on cognitive theories and frameworks such as mental workload~\cite{Zuger:2015}, level of task engagement~\cite{Pejovic:2015a}, or level of interruptibility~\cite{Yuan:2017}. NASA-TLX is a multi-dimensional rating tool which compromises six weighted scales including~\textit{mental}, \textit{physical}, and~\textit{temporal demand}, as well as~\textit{performance}, \textit{effort}, and~\textit{frustration}~\cite{Hart:1988}. However, the assessment of subjective workload using questionnaires may not be feasible in ubiquitous computing environments, where multiple tasks are performed simultaneously. Rather than using \textit{post hoc} questionnaires, implicit measures, including types of reaction, or time to react are used to evaluate attention management systems instead~\cite{Mehrotra:2015,Pielot:2014a}. For example, the authors in~\cite{Mehrotra:2015} measure the time taken to respond to a notification. If an individual responds within a period of $10$ minutes, the moment is classified as opportune and inopportune otherwise. The selection presented in Table~\ref{tab:models} is not meant to be a complete and comprehensive overview, given the fact that ubiquitous environments provide a variety of possibilities for modelling interruptibility. However, this overview points out that building interruptibility models in the field of attention management can benefit from interdisciplinary research, for example, by incorporating findings from psychology or cognitive science (see Section~\ref{sec:implications}).

\subsubsection{Managing Attention}
\label{sub:management}
Managing interruptions, once identified by the underlying algorithms and models, means that a system follows specific strategies to deal with interruptions (see Section~\ref{sub:interruptions}). In the past years, research focused on attention management systems which mediate and negotiate interruptions. Another strategy that has been used in attention management systems is the concept of \textit{mitigating} interruptions. Approaches following this strategy, for example, change the modality of a disruptive notification (e.g., vibration, silent, visual) rather than postpone them to an appropriate point in time. In Table~\ref{tab:management}, we list a few examples of applied management strategies in attention management systems. For example, Zulkernain et al. proposed an intelligent mobile interruption management system that utilizes a decision tree to pick an interruption mode (ring, silent, vibrate) based on the user's context automatically~\cite{Zulkernain:2010}. Rather than choosing alert types, other approaches focus on the utilization of Internet of Things (IoT) infrastructures~\cite{Weber:2016,Kubitza:2016}. The authors in~\cite{Kubitza:2016} propose a system that can track the location of an individual to deliver notifications to nearby output devices such as televisions, personal computers, or even ambient lights. Rather than deferring or changing the representation of interruptions, a few systems indicate the interruptibility status of individuals. For example, the system proposed in~\cite{Zuger:2017} indicates the interruptibility of employees via colored lights. Their findings show that employees perceive fewer interruptions over the day as they take the light status into account before interrupting colleagues.

However, an attention management system should adapt its handling strategy to a user's current needs. This means handling strategies, as well as underlying interruptibility models, should not be static and predefined beforehand. In this regard, choosing the~\textit{right} strategy remains one of the key challenges as an individual's behavior, and preferences might change over time. Different strategies and an adaptive mechanism that reacts to changes according to an individual's preferences are essential for such systems. One possible solution for adaptive attention management is proposed in~\cite{Mehrotra:2016a}. The authors introduce a mechanism that~\textit{mines} association rules by analyzing notification preferences of users using notification titles and different contextual information (e.g., location, activity, and time). Based on the user's action (dismiss or accept), the system generates a set of rules that reflect a user's attitude towards new notifications based on the current context. This approach is not only intuitive for the user, who consequently might put higher trust in an attention management system constructed in such a manner, but also interesting for further examination of theoretical frameworks that explain human interruptibility.

\begin{table}[ht]
	\caption{\textbf{Examples of attention management in ubiquitous computing environments.} In addition, approaches following the given management policies are listed.\label{tab:management}}{
		\begin{scriptsize}
			\centering
			\begin{tabular}{p{.1\columnwidth}p{.2\columnwidth}p{.375\columnwidth}}
				\toprule
				\textbf{Strategy} & \textbf{Management Policy} & \textbf{Approach} \\
				\toprule
				\textit{Mediating} 
				& Deferring to breakpoints & Physical activities~\cite{Okoshi:2015a,Okoshi:2015,Adamczyk:2004} \\
				& & Device interaction~\cite{Okoshi:2017,Poppinga:2014} \\
				& & Tasks \& Mental Workload~\cite{Iqbal:2007a,Bailey:2006,Iqbal:2008}
				\\ \midrule
				    
				\textit{Mitigating} & Choosing output device & Estimating device for receiving notifications~\cite{Weber:2016} \\
				& & Using IoT infrastructure for notifications delivery~\cite{Kubitza:2016} \\
				\cline{2-3}\addlinespace[0.5em] 
				& Choosing output modality & Inferring preferred output modalities~\cite{Zulkernain:2010}
				\\ \midrule
				
				\textit{Indicating}
				& Indicating interruptibility & Showing interruptibility status~\cite{Lai:2003,Begole:2004,Zuger:2017}
				\\ \bottomrule
			\end{tabular}
		\end{scriptsize}}
\end{table}

\subsection{Attention Management Systems Redefined}
Attention management encompasses more than balancing an individual's need for less disruptions and efficient information delivery. Attention management systems actively support individuals in managing interruptions and maintaining concentration on tasks and activities by constantly \textit{sensing}, \textit{modeling}, and \textit{managing} interruptibility. Models which help to determine interruptions, partly reflecting the actual state of an individual's interruptibility, represent a key factor of attention management systems. Based on our findings in Section~\ref{sec:ams}, we define attention management systems as
\begin{quote}
\textit{"systems that sense, model, and manage the attentional state of a user. Managing the attentional state is considered as any action of the system, which supports an individual to maintain its concentration on a task or activity."}
\end{quote}

\noindent\fbox{%
	\parbox{.98\textwidth}{%
		\begin{small}
			\textbf{Key factors of attention management systems}
			\begin{itemize}
				\item[$\star$] {Sensing, actuating, and machine learning are at the core of ubiquitous attention management systems. Sensing the physical environment allows for feature computation to discriminate concepts and factors of interruptibility.}
				\item[$\star$] {Interruptibility models hold assumptions and representations of an individual's interruptibility. Such models are based on features and contextual information that correlate with an individual's state of interruptibility.}
				\item[$\star$] {Based on the output of an interruptibility model, attention management systems follow different management strategies. Mediating and mitigating interruptions are the most commonly used strategies.}
			\end{itemize}

		\end{small}
	}%
}

\section{Implications and Perspectives}
\label{sec:implications}
Our attention is one of the most precious resources pervasive applications compete for. To protect individuals from adverse effects of attention steering, we have to ensure that future attention management systems respect ethical norms and work towards the well-being of their users. For the latter, it is crucial that attention management systems are designed in accordance with human cognitive processes and evaluated in long-term studies.

In this section, we discuss ethical and well-being implications of attention management. We highlight the lack of ethical guidelines when devising attention management systems and the absence of long-term studies on their impact on human well-being, and present suggestions for ameliorating this. We then identify untapped opportunities that arise from our review of cognitive psychology literature and identify key missing features that should be present in attention management systems.

\subsection{Attention Management vs. Ethics}
Two directions drive the evolution of attention management systems. On one side, we have a major part of academic research, based on Weiser's vision of~\textit{"calm technology"} that for the most part is invisible to the user~\cite{Weiser:1991}. On the other side, there are commercial applications that compete for attention, sending over $100$ notifications a day to individuals~\cite{Mehrotra:2015}. However, these two directions are not mutually exclusive. Commercial applications need to balance between grabbing a user's attention while avoiding to be excessively annoying. Similarly, \textit{"polite"} attention management systems still need to ensure that information is delivered before it becomes stale. Consequently, designers should ensure that attention management systems comply with the following ethical restrictions: \begin{enumerate*}[label=(\roman*)] \item{attention fragmentation incurred by attention-seeking applications is held at an acceptable level} and \item{time-critical information is delivered in a timely manner}\end{enumerate*}.

Satisfying the latter restriction requires a deep knowledge of an individual's engagement, the relationship of the to-be-conveyed information and the current, as well as, future tasks a person is engaged in. In practice, the sender, or the application, might tag notifications with their urgency level. In version 8.0 of their mobile operating system Android, Google has introduced~\textit{notification channels}, a paradigm that allows applications to expose different streams of notifications separately so that users can fine-tune an individual channel's appearance. While this addresses notification differentiation within a single app, system-level priority settings for notifications suffer from the misuse of individual applications that may set all of their notifications to the highest priority. Practical means of ensuring that condition (i) is satisfied have been examined through a large part of this article -- mobile sensing and machine learning harnessed to identify moments when a user is switching between tasks, is idle, or positively inclined towards the notification content, are the most commonly followed guidelines for designing \textit{"calm"} attention management systems.

Another important ethical concern is related to the direction in which attention is steered. Interruptions might be welcomed by the user; the sentiment might be positive, and also the time to react might be low, yet, the overall effect on an individual's well-being or productivity may be negative. This is particularly important in case the application is used by vulnerable groups, such as children or those who suffer from a mobile phone or gaming addiction. However, we should be careful to disentangle technological determinism from the way technology is used -- even when notifications are disabled on their phones, people still self-interrupt, checking for new content in fear of missing out~\cite{Oulasvirta:2012}.

\subsubsection*{Transparency. } To ensure that the manipulation of human attention does not affect an individual's autonomy to select, accept, or ignore information, attention management systems should be designed transparently. Individuals need to have the option to \textit{inspect}, \textit{override}, or to \textit{disable} functional principles. Interpretable attention management systems are one step in this direction, as they rely on human-readable rules to describe interruptibility preferences~\cite{Mehrotra:2016a,Mehrotra:2017}. A study by Mehrotra et al. indicates promising results towards transparent attention management systems, as more than $50\%$ of generated rules were accepted by individuals~\cite{Mehrotra:2017}. Advances at operating system levels, such as~\textit{notifications channels}, are one step towards transparent notification management.

\subsubsection*{Privacy. } Rich sensor data used to infer physical, physiological and mental states of individuals, information about the content of the message, relationship between the sender and receiver, have been widely used for efficient attention management. To ensure the privacy of both its users, as well as those interacting with users, attention management systems need to protect and safely process sensitive information. Such protection is especially important for systems that seek for a holistic view on an individual's interruptibility, gathering, exchanging, and storing information on multiple devices~\cite{Okoshi:2015a,Sykes:2014}. For systems that construct interruptibility models from data harvested from more than one individual, \textit{differential privacy} should be ensured, so that the probability of inferring individual behavior from data is minimized~\cite{Dwork:2006}. Finally, systems should incorporate best-practice privacy management approaches, such as the right for informational self-determination~\cite{Hornung:2009}.

\subsubsection*{Research questions. } Ethical concerns in ubiquitous attention management systems have not been widely addressed, yet. As guidelines for future efforts, we propose the following research questions:

\begin{quote}
\textit{(RQ1) How should the chain of ethical responsibility be established and who or what are the key factors in it?}
\end{quote}

\begin{quote}
\textit{(RQ2) How should transparent attention management systems be designed? How should functional principles of attention management systems be presented to the user?}
\end{quote}

\begin{quote}
\textit{(RQ3) What should the requirements for privacy-aware attention management systems be? Which privacy regulations should be considered, as they may vary among legislators or organizations?}
\end{quote}

\begin{quote}
\textit{(RQ4) How to ensure that legal requirements are integrated within the technical development process while designing attention management systems?}
\end{quote}

\subsection{Evaluating Effects of Attention Management Systems on Human Well-being}
Intelligent management of attention in ubiquitous systems may have a positive effect on human well-being~\cite{Schneider:2017,Zuger:2017,Iqbal:2008,Bailey:2006}. However, studies of such an effect are almost exclusively based on short time-scales, lasting approximately up to one month~\cite{Schneider:2017}. For example, Schneider et al. report short-term effects such as increased life balance, reduced stress levels, and reduced exhaustion of knowledge workers, after five-weeks of using mobile applications that manage availability~\cite{Schneider:2017}. Similar findings by Z{\"u}ger et al. suggest that, after two months of using their system, individuals feel less interrupted, more productive and experience more self-motivation~\cite{Zuger:2017}.

Although studies show positive short-term effects, long-term effects of attention management systems on human well-being and cognition are rather unknown. The reasons for the lack of longitudinal studies might be the following: \begin{enumerate*}[label=(\roman*)] \item{additional burden on the users coming from the introduction of a new technology}, \item{reluctance of users to change their routine because of an intervention}, and \item{the challenge of acquiring large groups of participants for longitudinal studies}\end{enumerate*}.
In fact, it has already been noted that the evaluation of attention management systems puts additional burden on individuals, as they have to be instructed to use new applications, fill out surveys regularly, or to be observed while using the evaluated solutions, which is the case if the \textit{shadowing} evaluation method is used~\cite{Gonzalez:2004,Mark:2005}. Also, most attention management systems are designed as~\textit{"black boxes"}. Without knowing functional principles, attention management restricts individuals' autonomy in selecting, accepting, and ignoring information. Individuals might resist the change these systems entail and try to restore settled routines~\cite{Miron:2006}. Finally, large groups of participants are required for reliable long-term evaluations. To the best of our knowledge, the largest pool in this domain, with more than 680,000 participants (albeit for only three weeks and not measuring users' well-being), was obtained by Okoshi et al. in collaboration with Yahoo! Japan~\cite{Okoshi:2017}. However, due to regulatory, privacy, and issues related to the additional burden imposed on participants, companies are in general reluctant to test prototype attention management systems' effect on well-being via production applications.

\subsubsection*{Towards Longitudinal Studies of the Effects of Attention Management Systems on Well-being. }
To facilitate long-term evaluation, we suggest that the following research directions are explored:
\begin{itemize}
    \item[-] {\textbf{Unobtrusive well-being detection.} Stress and annoyance are the most commonly reported negative effects of interruptions in ubiquitous computing environments~\cite{Mark:2008,Adamczyk:2004}. ESM is the most common means of assessing users' psychological well-being, within which standardized tests, such as NASA-TLX, or simple Likert scales for emotion inference are often used. However, ESM puts additional burdens on individuals and should be used moderately, which limits the amount of data that can be collected. Recently, a new wave of pervasive devices -- wearables -- including smartwatches, fitness wristbands, and smart glasses brought physiological sensing to everyday use. Since physiological signals are correlated with certain aspects of well-being, sensor data streams from wearable devices could be used for unobtrusive well-being inference. For example, Gjoreski et al. use wrist-worn photoplethysmogram (PPG) sensors, which measure heart rate and blood pulse volume, to infer the stress level of a user~\cite{Gjoreski:2016}. The inference, however, is not straightforward, as PPG readings are affected by hand movement. Thus, accelerometers are used to detect whether a user is mobile or not before the stress level inference is performed. Besides stress, affects~\cite{Shah:2015} and depression~\cite{Osmani:2015} can also be inferred unobtrusively thanks to mobile sensing.}
    
    \item[-] {\textbf{Specific considerations in evaluation study design.} While attention management systems provide a technical solution to control interruptions, non-technical environmental factors may play a significant role in how a user's attention is influenced. For example, studies indicate that the design of workplaces, the awareness towards interruptions, or organizational measures have a direct impact on the number of interruptions~\cite{Sykes:2011,Zuger:2017}. Therefore, efficient attention management systems need to consider such environmental factors. A holistic investigation of socio-technical aspects, including human, social, and organizational factors~\cite{Baxter:2011} might help define requirements for a long-term evaluation for a given environment. Without socio-technical requirements, individuals might perceive attention management systems as non-supportive, too complex to use~\cite{Schneider:2017}, or even quit studies if they feel that a system does not reflect their interruptibility~\cite{Zuger:2017}. In one of our previous studies, users were reluctant to complete their study participation as the developed attention management tool did not support their well-established routine of using Skype for business communication~\cite{Schneider:2017}. Although, we focus on well-being in this section, the above concerns are relevant for all long-term evaluation studies of attention management systems.}
\end{itemize}

We conclude the discussion with two questions meriting research:
\begin{quote}
\textit{(RQ6) How should a long-term evaluation of the well-being impact of attention management systems be performed provided minimal user burden and a large number of participants?} The above discussion on unobtrusive well-being state detection could be a starting point for further exploration. Particularly, since physiological data can be incorporated in attention inference algorithms~\cite{Haapalainen:2010,Zuger:2015}.
\end{quote}

\begin{quote}
\textit{(RQ7) How to incorporate contextual, social, and organizational factors when planning long-term evaluation studies?} These aspects are not only important for evaluations but could also give essential features to the design of attention management systems.
\end{quote}

\subsection{Towards Theoretically-Grounded Attention Management Systems}
\label{sub:challenges_ams}
Research on cognitive processes and ubiquitous computing systems practice have evolved separately, and one of the main goals of this survey is to identify untapped opportunities for building more efficient attention management systems that take into account the underlying psychological processes. In this subsection, we point out these potential research avenues, briefly discuss their theoretical underpinnings, and present work that, at least partly, addresses them.

\subsubsection*{Interrupting at task boundaries or idle times. } Times, when a user is not actively engaged in a task, are the most suitable moments for interruption according to both theoretical frameworks~\cite{Miyata:1986}, as well as experimental studies~\cite{Adamczyk:2004,Iqbal:2005}. From the memory for problem state theory point of view, at such moments the problem state, i.e., the intermediate mental representation necessary for performing a task, is not needed. Thus there is no interruption along procedural or declarative memory resources. Furthermore, task engagement flows through different stages: planning, execution, and evaluation~\cite{Cutrell:2000}. In a study of instant messaging users, Czerwinski et al. found that the participants took more time to attend an interruption in case it arrived during the execution phase than in the other two phases~\cite{Czerwinski:2000}. Interestingly, they also find a disruptive effect of an interruption in the evaluation phase. The authors hypothesize that this may reflect the time required for users to visually re-orient themselves to where they left off and concluded the task completion. Attention management systems, such as Attelia II~\cite{Okoshi:2015a}, successfully apply defer-to-breakpoint policies~\cite{Iqbal:2007a} within physical activities. Inferring one's mental engagement, on the other hand, is not straightforward. Certain physiological signals, such as pupil dilation, correlate with mental load, and indeed have been used for assessing task engagement~\cite{Haapalainen:2010}. However, a reliable means of detecting a user's task engagement using commodity devices is yet to be demonstrated~\cite{Urh:2016}. Finally, Pielot et al. demonstrate that bored users are likely to engage with recommended content~\cite{Pielot:2015,Pielot:2015a}.

\begin{quote}
\textit{(RQ8) How can we use commodity pervasive computing devices, such as wearable sensors, to infer a user's task engagement and consequently enhance attention management systems?}
\end{quote}

\subsubsection*{Problem state rehearsal. } Bluma Zeigarnik's experiments from 1927 showed that people remember uncompleted or interrupted tasks better than completed tasks~\cite{Zeigarnik1927}. Therefore, there is a natural tendency to return to the primary task when interrupted. However, the complexity of the original task plays a significant role in the success of such a switchback. According to the Salvucci and Taatgen's theory, complex tasks are difficult to return to because they require the retrieval of the problem state from declarative memory. The more time the problem state spends in the declarative memory, the more likely it is that the retrieval will fail due to the decay of the stored information~\cite{Salvucci:2010a}. Borst et al. argue that the rehearsal of the problem state should come~\textit{before} a user switches to the secondary task~\cite{Borst:2010}. Along the same lines, Cades et al. found that interruptions are more disruptive when they minimized the participant's ability to rehearse the primary task during the interruption, not necessarily when the secondary task is more difficult~\cite{Cades:2007}. Trafton et al. have conducted experiments to demonstrate that visual alerts before a forced switch to another task help users rehearse the primary task state, and consequently facilitate the later switchback~\cite{Trafton2003}. Currently, the rehearsal of the problem state represents an untapped challenge for practical attention management systems. Thus, we propose the following questions for future research:

\begin{quote}
\textit{(RQ9) How to condense and prepare the primary task for problem state rehearsal automatically? How to represent the content to an individual, so that the problem state is captured at a glance?}
\end{quote}

\begin{quote}
\textit{(RQ10) How long should the problem state rehearsal last, so that an interruption does not get stale?}
\end{quote}

\subsubsection*{Providing hints to revert the context. } To facilitate a switchback, Salvucci and Trafton's theory suggest that cues related to the old problem state should be provided. Altmann and Trafton have experimentally confirmed that neighboring elements of the primary task are better retrieval cues than elements separated by greater psychological distance~\cite{Altmann2007}. HCI researchers have been tackling this issue since the rise of multitasking PC applications. Czerwinski et al. have conducted a diary study of PC users to identify possible design improvements to facilitate task switchbacks. The authors propose time-centric visualizations and tools that reconfigure the layout of desktop windows (i.e., content and applications) that compromise a task~\cite{Czerwinski:2004}. Also, Iqbal and Horvitz examined the role of visual hints on task retrieval and found that the amount of visual obfuscation caused by an interruption impacts the time to return to the primary task -- interruptions that cover only a small portion of a desktop are easier to return from~\cite{Iqbal:2007}. Following this argumentation, attention management systems need to construct and provide psychologically close hints to support switchbacks to primary tasks. In~\cite{Srinivas:2016}, the authors design a system for managing interruptions in intensive care units. They argue that color-coded notifications act as psychological cues that can be used to retrieve information about the patient it concerns. In this regard, the color of the notification helps to switchback to the patient~\textit{at a glance}. However, the automatic construction of psychological hints is still in its origin, and the following questions might help to steer future research in this area:

\begin{quote}
\textit{(RQ11) What are psychologically close hints and how can they be constructed automatically?}
\end{quote}

\begin{quote}
\textit{(RQ12) If multiple tasks can be resumed after an interruption, which hint should be displayed?}
\end{quote}

\subsubsection*{Relevance and relation to the primary task. } Findings indicate that the interruption content plays an important role in the perceived disruptiveness of interruptions~\cite{Speier:1999}. In fact, interruptions that contain relevant information for the primary task may negatively affect the time to complete the task but may help with the task performance quality. Interruptions with content irrelevant to the primary task also increase completion time but lead to higher task error rates~\cite{Addas:2015}. A similar approach to relevance and relation of interruptions to the primary task is presented in~\cite{Anderson:2016}. The authors assume that an interruption is more relevant to individuals if it is matched with their social role (e.g., employee, family member). Consequently, attention management systems should incorporate the relevance of interruptions to the primary task. In this regard, we propose the following question:

\begin{quote}
\textit{(RQ13) How can an attention management system infer the relevance of an interruption to a task at hand, and at how should interruptions with different levels of similarity with the primary task be handled?}
\end{quote}

\subsubsection*{Multimodal alerts and notification representations. } In ubiquitous systems, interruptions are often signaled through notifications that may use different alert types -- sound, light, or vibration. Also, incoming information can be conveyed using different means -- written on a screen, spoken through speakers, or by using a buzz alarm. The alert and message representation types may not align well with a user's current task engagement. According to the cognitive theories presented in Section~\ref{sec:theory} different cognitive resources could be occupied at a time. If a user is engaged in a visual task only, perhaps an aural signal and notification whose contents are read to the user are still appropriate. An example would be a driving situation, where an incoming text message is less likely to distract the driver if read out aloud without requiring the driver to divert visual attention from the road. Consequently, a research question arises:

\begin{quote}
\textit{(RQ14) Can ubiquitous technologies infer which of the user's cognitive resources are occupied and change the alert and information representation modality in a way to minimize disruption?}
\end{quote}

\section{Summary}
\label{sec:summary}
In this article, we discussed the emerging field of attention management in ubiquitous environments. Particular attention has been paid to underlying theoretical frameworks and concepts of practical attention management systems as well as on actual implementations and technical designs. By leveraging unique characteristics of ubiquitous computing, e.g., personalized information, always-on connectivity or sensing capabilities, combined with cognitive and psychological concepts, attention management systems can help mitigate and mediate the ever-increasing number of interruptions. Fueled by theoretical and practical findings on the nature of human interruptibility, we discussed design implications for attention management systems, e.g., interrupting at task boundaries or providing cues for task retrieval. Actual implementations of such designs in the form of interruptibility models embedded in applications rely on mobile sensing, actuation, and machine learning. Although essential to efficiently support individuals in their attention management, building interruptibility models remains a challenge in the field of attention management. Identifying correlations between inferred contextual descriptors and an individual's actual state of interruptibility represents an open research question. In this regard, we showed that attention management systems could benefit from interdisciplinary research as the facets of human interruptibility concern various research field, e.g., cognitive science, human-computer-interfaces, or psychology. In fact, certain existing interruptibility models already incorporate findings from adjacent research fields, for example, interrupting at low cognitive states or between transitions of physical activities.

The capability to continuously sense, process, model, and to manage interruptibility facilitate adaptive attention management, more specifically, systems that can~\textit{sense} and~\textit{understand} the user. In the next few years, attention management will leverage existing and future modalities, e.g., smart clothes, sophisticated physiological measures, allowing for more efficient interruptibility models. Further investigations and efforts in this interesting field will help to shape and form efficient and advanced attention management systems.

\bibliographystyle{ACM-Reference-Format}
\bibliography{bibliography}
\end{document}